\documentclass[Times,4pt,aps,pra,twocolumn,showpacs,amsmath,amssymb,floatfix,footinbib,superscriptaddress]{revtex4-2}
\usepackage{amsmath,natbib,graphicx,subfigure,amssymb,graphics,amsmath,mathrsfs,CJK,color,comment}
\usepackage{multirow,fancyhdr,color,bm,tabularx,psfrag,geometry,dcolumn,datetime,physics,booktabs}
\usepackage[colorlinks=true,linkcolor=blue,urlcolor=blue,citecolor=blue]{hyperref}
\usepackage[mathlines]{lineno}
\def\footnoterule{\kern -1mm \hrule width 5.8cm \kern 2.2mm}
\usepackage{tikz,xcolor,hyperref}
\definecolor{lime}{HTML}{A6CE39}
\DeclareRobustCommand{\orcidicon}{%
    \begin{tikzpicture}
    \draw[lime, fill=lime] (0,0)
    circle [radius=0.16]
    node[white] {{\fontfamily{qag}\selectfont \tiny ID}};\draw[white, fill=white] (-0.0625,0.095)
    circle [radius=0.007];
    \end{tikzpicture}
    \hspace{-2mm}}
\foreach \x in {A, ..., Z}
{\expandafter\xdef\csname orcid\x\endcsname{\noexpand\href{https://orcid.org/\csname orcidauthor\x\endcsname}{\noexpand\orcidicon}}}

\begin{document}
\title{Coherent catalyst induced stabilization of ergotropy in open quantum batteries}

\author{Ni-Ya Zhuang}
\affiliation{Center for Quantum Materials and Computational Condensed Matter Physics, Faculty of Science, Kunming University of Science and Technology, Kunming, 650500, PR China}
\author{Shun-Cai Zhao\orcidA{}}
\email[Corresponding author: ]{zsczhao@126.com.}
\affiliation{Center for Quantum Materials and Computational Condensed Matter Physics, Faculty of Science, Kunming University of Science and Technology, Kunming, 650500, PR China}

\begin{abstract}
Environmental dissipation and thermal fluctuations fundamentally constrain the extractable work and long-time stability of open quantum batteries. To mitigate dissipation-induced energy degradation without external driving protocols, we propose a cavity-mediated hybrid quantum battery coupled to an auxiliary coherent qubit. Using the Lindblad master equation and ergotropy analysis, we show that coherent interference between different interaction channels generates a decoherence-free-like invariant subspace that suppresses relaxation-induced energy leakage and stabilizes the steady-state ergotropy. The resulting protection mechanism remains effective under strong dissipation and finite-temperature conditions, indicating that interference-assisted coherent control may provide a feasible strategy for robust quantum energy storage in nonequilibrium open systems.
\end{abstract}
\maketitle

\section{Introduction}
The pursuit of high-performance quantum batteries (QBs) has become a focal point in quantum thermodynamics\cite{PhysRevLett.129.130602,PhysRevLett.122.047702,PhysRevLett.120.117702}, as these devices promise to revolutionize energy storage by exploiting quantum resources such as entanglement and coherence to achieve superior charging power~\cite{PhysRevE.102.052109,PhysRevE.87.042123}. In recent years, research efforts in this field have mainly concentrated on three interconnected directions: first, the design of many-body battery architectures, such as spin chains and Hubbard models~\cite{PhysRevA.103.033715,PhysRevResearch.2.023095,PhysRevB.109.165121}, to enhance charging power and storage capacity; second, the investigation of open-system effects, particularly how environmental decoherence and dissipation deteriorate the stored ergotropy~\cite{PhysRevB.108.125420,PhysRevA.102.060201,PhysRevA.105.012212}; and third, the development of protection strategies, including high-$Q$ cavities, structured reservoirs inducing non-Markovian backflow, decoherence-free subspaces, and dynamical decoupling sequences~\cite{PhysRevLett.81.2594, HADIPOUR2024107928, Li:22, Hadipour2025, PhysRevB.99.205437}.

While these pioneering works have significantly advanced our understanding of open quantum batteries, each approach carries inherent limitations that become critical in realistic dissipative environments. For instance, passive isolation using high-$Q$ cavities or engineered reservoirs can suppress dissipation\cite{PhysRevLett.107.010402,PhysRevA.112.043705,b5lx-j66h,43n6-rnj3,Liu2025Dissipation,p93y-jflt}, but it often demands extreme physical parameters (e.g., ultralow temperatures or highly precise spectral shaping) that are difficult to scale or maintain in practical architectures\cite{PhysRevLett.131.196602,PhysRevResearch.6.043306,PhysRevLett.120.123601}. Active dynamical control methods, such as external driving fields or decoherence-free subspaces, offer more flexibility\cite{PhysRevLett.124.110603,PhysRevA.65.042317,ndlt-qszr,PhysRevE.105.044125,PhysRevA.107.032203}; however, they typically rely on fast, time-dependent pulses or specific symmetry conditions, which may introduce additional noise channels and are experimentally demanding\cite{PhysRevE.105.044125,PhysRevA.107.032203}. More importantly, most existing proposals either assume a globally uniform environment or treat the battery as a closed system with weak perturbations\cite{PhysRevA.98.053862,PhysRevA.100.042322,PhysRevLett.124.140502}, thereby neglecting the possibility of actively reshaping the local relaxation landscape using auxiliary coherent units\cite{PhysRevLett.104.190401,PhysRevE.103.042118}. This omission can lead to overoptimistic predictions of energy retention, especially under strong cavity loss and finite-temperature conditions-a regime where practical quantum batteries must ultimately operate\cite{PhysRevE.109.014131,ndlt-qszr,PhysRevLett.134.180401,PhysRevLett.121.123601}.

Building upon previous studies on open quantum system dynamics~\cite{56-PENG2026131462,49-10.1140/epjp/s13360-023-04660-4,53-xqtv-qbyk}, we investigate robust energy storage in many-body quantum batteries under dissipative conditions. To overcome the limitations of passive protection and external control protocols, we introduce an auxiliary coherence-assisted qubit that actively participates in the system dynamics. The core mechanism of this work relies on destructive quantum interference between cavity-spin and catalyst-spin interaction channels, which introduces an extra tunable parameter to actively reshape the effective relaxation pathways of the open quantum battery. Importantly, the auxiliary catalytic qubit preserves its total energy throughout the entire dynamical evolution, behaving as a true energy-conserving catalyst~\cite{RevModPhys.96.025005,PhysRevA.107.042419} that modifies dissipative channels without net energy exchange or consumption. This interference-based mechanism enables a continuous and autonomous modulation of dissipation effects without requiring external pulsed driving and is compatible with current circuit-QED and trapped-ion platforms~\cite{PhysRevE.109.054132,PhysRevLett.134.180401,PhysRevResearch.6.023136}. Our results suggest that such a catalytic design offers a practical route toward enhancing energy retention and stabilizing quantum batteries in noisy environments.

The remainder of this paper is organized as follows: In Sec.~II, we describe the theoretical model of the spin-chain battery, the catalytic qubit, and the cavity-mediated dissipative environment, along with the master equation formalism and the figure of merit (ergotropy). Sec.~III presents numerical analysis of energy storage performance under various dissipative regimes, comparing the catalyst-assisted scheme with conventional unprotected configurations. In Sec.~IV, we discuss the underlying physical mechanisms-in particular the role of coherent interference in creating a protected quasi-dark state-and conclude with a summary of the implications for future experimental implementations.

\section{Model and Theoretical Formalism}

Practical realization of quantum batteries is fundamentally challenged by environmental decoherence, which typically leads to the rapid depletion of stored energy~\cite{d9k1-75d4,PhysRevLett.132.090401,PhysRevResearch.2.013095}. Previous efforts to mitigate this issue have largely focused on passive protection strategies~\cite{PhysRevLett.134.180401,PhysRevApplied.14.024092}, such as exploiting dark states of the system-environment interaction or employing dynamical decoupling sequences. While these approaches have demonstrated some success in prolonging coherence, they often suffer from notable drawbacks: dark-state-based schemes typically require precise parameter matching and offer limited tunability, whereas dynamical decoupling relies on fast, time-dependent control pulses that are experimentally demanding and may introduce additional noise channels~\cite{PhysRevA.107.032615,PhysRevB.111.064503,PhysRevA.111.052622}. Moreover, most existing proposals~\cite{cyrc-ms34,4hr3-rl2y,msh1-b1cc} assume either a globally uniform environment or a simple Markovian bath, neglecting the possibility of actively reshaping the local relaxation landscape using auxiliary coherent units.

\begin{figure}[htbp]
  \centering
  \includegraphics[width=0.8\linewidth]{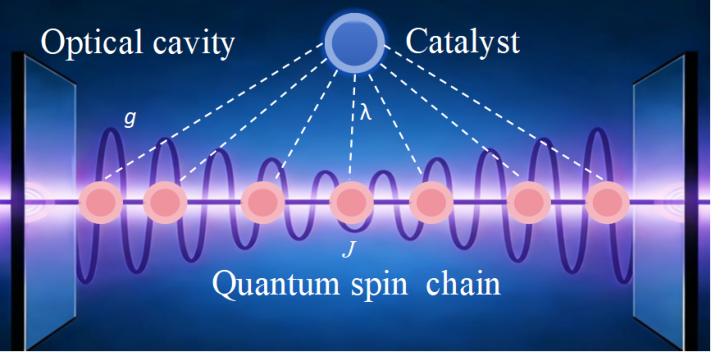}
  \caption{Schematic of the catalyst-assisted quantum spin-chain battery.}
  \label{fig:fig1}
\end{figure}

To overcome these limitations, we propose a composite quantum battery architecture that incorporates a coherent catalytic unit to actively modulate the relaxation landscape. As schematically illustrated in Fig.~\eqref{fig:fig1}, the system consists of an $N$-site quantum spin chain~\cite{PhysRevLett.120.117702} acting as the energy storage medium, integrated within a single-mode lossy optical cavity. The catalyst is a coherence-assisted qubit that couples directly to each spin in the chain. By tuning the catalyst-battery coupling strength and relative phase with respect to the cavity-battery interaction, we engineer a coherent interference between the two coupling channels. This interference creates a spatially inhomogeneous intra-cavity field profile, characterized by a localized suppression of the field amplitude near the spin chain.

Under the rotating-wave approximation (RWA)~\cite{PhysRevA.86.023824,PhysRevLett.99.173601}, the total Hamiltonian of the coupled system is written as $\hat{H} = \hat{H}_0 + \hat{H}_{\text{int}}$, with $\hbar=1$. The free Hamiltonian $\hat{H}_0$ accounts for the uncoupled energies of the cavity field, the spin-chain battery, and the catalytic qubit:
\begin{equation}\label{1}
\hat{H}_0 = \Omega_c \hat{a}^\dagger \hat{a} + \Omega_a \sum\limits_{i=1}^{N} \hat{\sigma}_i^\dagger \hat{\sigma}_i + \frac{\Omega_{\text{cat}}}{2} \hat{\sigma}_z^{(\text{cat})},
\end{equation}

\noindent where $\hat{a}^\dagger$ ($\hat{a}$) are the creation (annihilation) operators for the cavity mode with frequency $\Omega_c$, and $\hat{\sigma}_i^\dagger$ ($\hat{\sigma}_i$) are the raising (lowering) operators for the $i$-th spin with transition frequency $\Omega_a$. The interaction Hamiltonian $\hat{H}_{\text{int}}$ describes the nearest-neighbor exchange within the battery, the cavity-spin coupling, and the catalyst-spin coupling:
\begin{eqnarray}\label{2}
\hat{H}_{\text{int}} &=& J \sum\limits_{i=1}^{N-1} \left( \hat{\sigma}_i^\dagger \hat{\sigma}_{i+1} + \text{H.c.} \right) + g \sum\limits_{i=1}^{N} \left( \hat{a}^\dagger \hat{\sigma}_i + \text{H.c.} \right) \nonumber\\
&& + \lambda \sum\limits_{i=1}^{N} \left( \hat{\sigma}_+^{(\text{cat})} \hat{\sigma}_i + \text{H.c.} \right).
\end{eqnarray}

Here, $J$ is the nearest-neighbor exchange coupling, $g$ is the cavity-spin coupling strength, and $\lambda$ is the catalyst-spin coupling strength. The relative sign between $g$ and $\lambda$ (implicit in the Hamiltonian) controls the interference between the two interaction channels.

To simulate the open-system evolution of the density matrix $\hat{\rho}(t)$ under cavity loss and thermal fluctuations, we employ the Lindblad master equation~\cite{Lindblad1976}:
\begin{equation}\label{3}
\frac{d\hat{\rho}(t)}{dt} = -i[\hat{H}, \hat{\rho}(t)] + \kappa \mathcal{D}[\hat{a}]\hat{\rho}(t) + \sum_i \gamma_i \mathcal{D}[\hat{\sigma}_i]\hat{\rho}(t).
\end{equation}

Here $\kappa$ is the cavity decay rate, and $\gamma_i$ is the relaxation rate of the $i$-th spin mode induced by the thermal reservoir. The Lindblad dissipators are defined as
    \[
    \mathcal{D}[\hat{a}]\hat{\rho} = (n+1)\hat{a}\hat{\rho}\hat{a}^\dagger - \frac{1}{2}n\big\{\hat{a}^\dagger \hat{a}, \hat{\rho}\big\}
    \]
    and
    \[
    \mathcal{D}[\hat{\sigma}_i]\hat{\rho} = \hat{\sigma}_i\hat{\rho}\hat{\sigma}_i^\dagger - \frac{1}{2}\big\{\hat{\sigma}_i^\dagger \hat{\sigma}_i, \hat{\rho}\big\},
    \]
\noindent where the thermal occupation number $n$ is given by the Bose-Einstein distribution
\begin{equation}\label{4}
n = \frac{1}{e^{\Omega_c / (k_B T)} - 1},
\end{equation}

\noindent with $\Omega_c$ the cavity resonant frequency, $k_B$ the Boltzmann constant, and $T$ the temperature of the environment.

The performance of the battery is quantified by the ergotropy $W_{\text{ext}}(t)$, which represents the maximum extractable work from the battery subsystem~\cite{Allahverdyan2004}:
\begin{equation}\label{5}
W_{\text{ext}}(t) = \text{Tr}\left[\hat{\rho}(t)\hat{H}_{\text{battery}}\right] - \min_{\hat{U}} \text{Tr}\left[\hat{U}\hat{\rho}(t)\hat{U}^\dagger \hat{H}_{\text{battery}}\right],
\end{equation}

\noindent where the minimization is over all cyclic unitary operations $\hat{U}$ on the battery. This measure captures the useful portion of stored energy that can be coherently extracted~\cite{56-PENG2026131462}. Using this framework, we will analyze the long-time dynamics of $W_{\text{ext}}(t)$ to assess whether the proposed catalytic interference mechanism can protect battery performance against decoherence and energy leakage.

\section{Results and Numerical Verification}

We now analyze the dynamical performance of the catalyst-assisted quantum battery. To highlight the role of catalytic coherent interference, we compare two representative scenarios: (a) a catalyst-free configuration, where the battery interacts only with the cavity mode; and (b) the full catalyst-assisted setup. The parameters used in each case are summarized in Table~\ref{tab:params}.

\begin{table}[htbp] 
  \centering
  \caption{Simulation parameters for catalyst-free (a) and catalyst-assisted (b) QBs. Spin number $N$=$3$; ``$/$'' denotes the scanned parameter.}\vspace{-2em}
  \setlength{\tabcolsep}{0.2cm}\vspace{2em}
  \begin{tabular}{|l|c|c|c|c|c|c|c|c|}
    \hline
     & \multicolumn{4}{c|}{(a)($\lambda = 0$)} & \multicolumn{4}{c|}{(b)($\lambda = 1.5$)} \\
    \cline{2-9}
            & $g$   &  $J$  & $\kappa$   & $T$  & $g$  &  $J$  &$\kappa$  &   $T$  \\
    \hline 
    Fig.\,2 & 0.3   &   /   &   0.15     & 0.8  & 0.3  &    /  &   0.15   &   0.8    \\
     \hline
    Fig.\,3 & /     &  1.6  &   0.15     & 0.8  & /    &  1.6  &   0.15   &    0.8   \\
     \hline
    Fig.\,4 & 0.3   &  1.6  &    /       & 0.8  & 0.3  &  1.6  &   /      &    0.8   \\
     \hline
    Fig.\,5 & 0.3   &  1.6  &   0.15     &  /   & 0.3  &  1.6  &   0.15   &    /    \\
    \hline
  \end{tabular}
\label{tab:params}
\end{table}

\begin{figure}[!htb]
  \centering
  \includegraphics[width=0.49\linewidth]{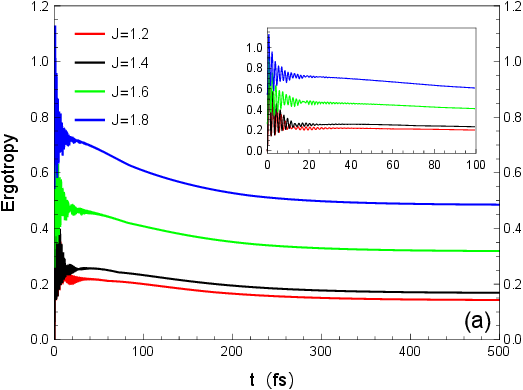}  \includegraphics[width=0.49\linewidth]{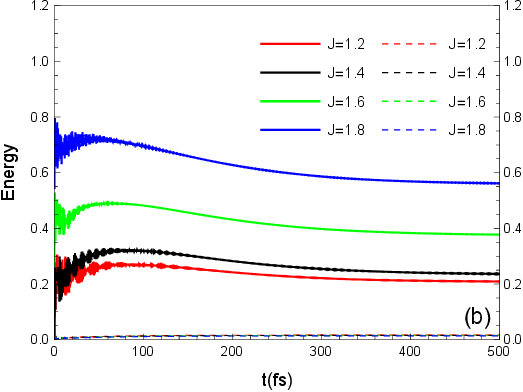}
  \caption{(a) Ergotropy dynamics of quantum batteries under different internal coupling strengths $J$. (b) Energy dynamics of the batteries (solid lines) and the catalyst (dashed lines) with varying  $J$. $\Omega_c$=0.5,  $\Omega_a$=2.0,  $\Omega_{cat}$=0.06, $k_{B}$=1, and other parameters are listed in Tab.~\eqref{tab:params}.}
  \label{fig2}
\end{figure}

\begin{figure}[!htb]
  \centering
 \includegraphics[width=0.49\linewidth]{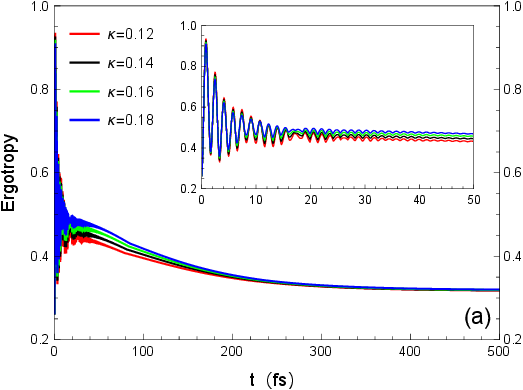}  \includegraphics[width=0.49\linewidth]{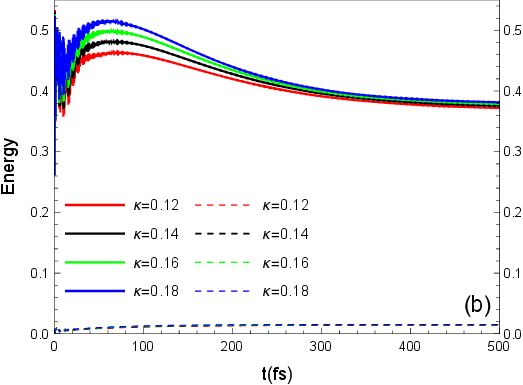}
  \caption{(a) Ergotropy dynamics of quantum batteries under different cavity decay rate $\kappa$. (b) Energy dynamics of the batteries (solid lines) and the catalyst (dashed lines) with varying  $\kappa$. Other parameters are the same to those in Fig.~\ref{fig2}.}
  \label{fig3}
\end{figure}

\begin{figure}[!htb]
  \centering
 \includegraphics[width=0.49\linewidth]{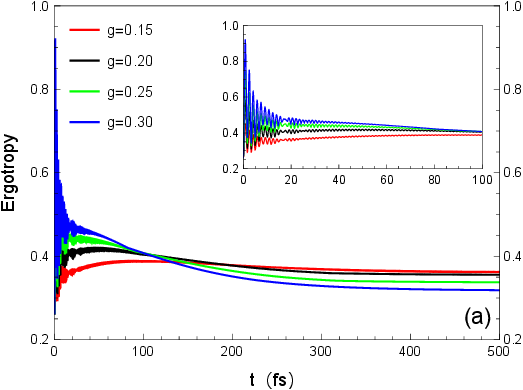}  \includegraphics[width=0.49\linewidth]{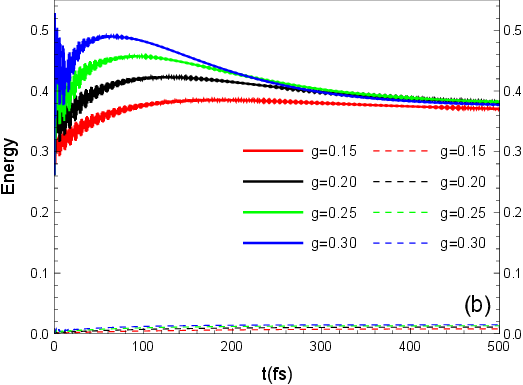}
  \caption{(a) Ergotropy dynamics of quantum batteries under different cavity-spin coupling strengths $g$. (b) Energy dynamics of the batteries (solid lines) and the catalyst (dashed lines) with varying  $g$. Other parameters are the same as those in Fig.~\ref{fig2}.}
  \label{fig4}
\end{figure}

\begin{figure}[!htb]
  \centering
 \includegraphics[width=0.49\linewidth]{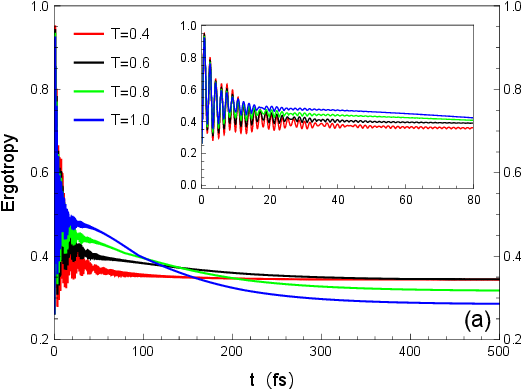}  \includegraphics[width=0.49\linewidth]{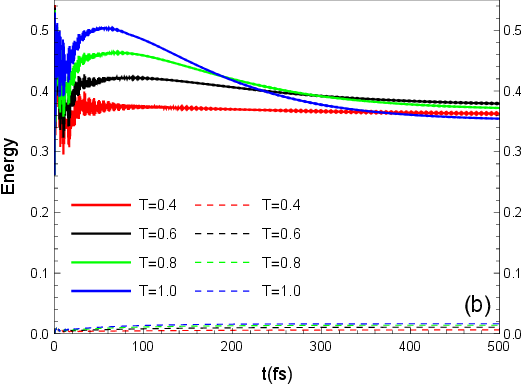}
  \caption{(a) Ergotropy dynamics of quantum batteries under different environment temperature $T$. (b) Energy dynamics of the batteries (solid lines) and the catalyst (dashed lines) with varying  $T$. Other parameters are the same as those in Fig.~\ref{fig2}.}
  \label{fig5}
\end{figure}

Figs.~\ref{fig2}--\ref{fig5} present the dynamical evolution of ergotropy and stored energy for the two scenarios. In each figure, scenario (a) (catalyst-free) shows the battery performance under independent modulation of internal coupling, cavity decay, cavity-spin coupling, or temperature, with insets highlighting short-time transient behavior. Scenario (b) (catalyst-assisted) displays the battery energy (solid lines) together with the energy of the catalytic qubit (dashed lines). The flat dashed lines in Figs.~\ref{fig2}--\ref{fig5} confirm the constant-energy characteristic of the coherent catalyst, consistent with the notion of an energy-conserving catalyst~\cite{RevModPhys.96.025005,PhysRevA.107.042419}.

We first examine the effect of internal spin-exchange coupling $J$ (Fig.~\ref{fig2}). In the catalyst-free case [Fig.~\ref{fig2}(a)], increasing $J$ enhances the ergotropy, but noticeable transient oscillations persist. Introducing the catalyst [Fig.~\ref{fig2}(b)] not only suppresses these initial oscillations but also raises the steady-state ergotropy. For instance, at $J=1.8$ and $t=500$ fs, the ergotropy increases from $\approx 0.48$ in the bare system to above $0.56$ with the catalyst.

Fig.~\ref{fig3} addresses the impact of cavity decay $\kappa$. The catalyst-free case shows strong sensitivity to $\kappa$, with ergotropy decaying more rapidly at larger $\kappa$ [Fig.~\ref{fig3}(a)]. The catalyst-assisted configuration [Fig.~\ref{fig3}(b)] markedly suppresses the initial transient oscillations and yields a modest but clear increase in steady-state ergotropy as $\kappa$ rises. This behavior is particularly evident for $\kappa=0.18$ (blue curves), demonstrating that the catalyst effectively counteracts cavity-induced dissipation.

A more striking demonstration of catalytic protection appears in Fig.~\ref{fig4}, where we vary the cavity-spin coupling strength $g$. In the bare system [Fig.~\ref{fig4}(a)], the steady-state ergotropy decreases with increasing $g$, indicating that stronger cavity-battery coupling exacerbates energy leakage. Remarkably, this adverse trend is reversed by the catalyst [Fig.~\ref{fig4}(b)]: even under strong coupling ($g=0.3$), the catalyst-assisted battery maintains a steady-state ergotropy significantly higher than that of the bare system at $g=0.15$. This reversal directly evidences the decoherence-protection mechanism enabled by coherent interference between cavity-mediated and catalyst-mediated channels.

Finally, Fig.~\ref{fig5} examines the effect of environmental temperature $T$. In the catalyst-free scenario [Fig.~\ref{fig5}(a)], increasing $T$ reduces the steady-state ergotropy, reflecting thermal decoherence. The catalyst-assisted system [Fig.~\ref{fig5}(b)] substantially mitigates this degradation: at $T=1.0$, the steady-state ergotropy rises from approximately $0.28$ to about $0.35$. Collectively, these results demonstrate that the catalytic coherent interference mechanism provides robust, autonomous protection against diverse dissipation channels, including internal decoherence, cavity loss, and thermal noise.

\section{Physical Mechanism: Liouvillian Spectral Restructuring and Catalytic Quasi-Dark-State Protection}
\label{sec:mechanism}

To clarify the microscopic origin of the enhanced ergotropy retention and the suppression of dissipation observed in Figs.~\ref{fig2}--\ref{fig5}, we analyze the open-system dynamics from the perspective of Liouvillian spectral restructuring. The evolution of the density matrix is governed by the Liouvillian superoperator $\mathcal{L}$, defined via $\dot{\hat{\rho}} = \mathcal{L}\hat{\rho}$. The real parts of its complex eigenvalues, $\operatorname{Re}(\Lambda_k)$, dictate the relaxation timescales of the corresponding dynamical modes. Consequently, the long-time preservation of work-extraction potential is fundamentally determined by the emergence of slowly decaying, metastable Liouvillian eigenstates.

In the bare battery configuration ($\lambda$=0), the many-body spin chain collectively couples to the lossy cavity mode via the light-matter interaction $\hat{H}_{\mathrm{cav}} $=$g\sum\limits_{i=1}^N (\hat{a}^\dagger\hat{\sigma}_i^- + \text{H.c.})$. Because the cavity mirror induces photon loss at a rate $\kappa$, the collective spin excitations strongly hybridize with the rapidly decaying photonic degrees of freedom. This hybridization shifts the relevant Liouvillian branches deep into the left half of the complex plane ($\operatorname{Re}(\Lambda_k) \ll 0$), leading to a catastrophic washout of stored ergotropy, as confirmed by the bare dynamics in Figs.~\ref{fig2}(a) and \ref{fig3}(a). Crucially, scaling up the coupling strength $g$ under these conditions merely accentuates this detrimental hybridization, accelerating the energy leakage [see Fig.~\ref{fig4}(a)].

The introduction of the auxiliary coherent qubit counters this decay by establishing a competing, energy-conserving interaction channel $\hat{H}_{\mathrm{cat}} $=$ \lambda \sum\limits_{i=1}^N (\hat{\sigma}_+^{(\text{cat})}\hat{\sigma}_i^- + \text{H.c.})$. In the uniform coupling limit, we can naturally define the collective spin lowering operator as $\hat{S}^- $=$ \frac{1}{\sqrt{N}}\sum\limits_{i=1}^N \hat{\sigma}_i^-$, which allows us to cast the total interaction Hamiltonian $\hat{H}_{\mathrm{int}} $=$ \hat{H}_{\mathrm{cav}} + \hat{H}_{\mathrm{cat}}$ into a compact collective form:
\begin{equation}
\hat{H}_{\mathrm{int}} = \sqrt{N}g \bigl(\hat{a}^\dagger\hat{S}^- + \hat{a}\hat{S}^+\bigr) + \sqrt{N}\lambda \bigl(\hat{\sigma}_+^{(\text{cat})}\hat{S}^- + \hat{\sigma}_-^{(\text{cat})}\hat{S}^+\bigr).
\end{equation}

\noindent When the control parameters are tuned to satisfy the phase-matching and amplitude-matching interference condition,
\begin{equation}
\lambda \approx -\frac{g}{\sqrt{N}},
\label{eq:condition}
\end{equation}

\noindent the cavity-mediated and qubit-mediated transition amplitudes interfere destructively within the single-excitation manifold. This quantum cancellation effectively decouples the collective atomic excitation from the lossy cavity mode. In Liouville space, this spectral restructuring manifests as the isolation of a long-lived, metastable eigenmode characterized by $\operatorname{Re}(\Lambda_{\mathrm{meta}}) \rightarrow 0$. At the state level, this dark mode corresponds to a collective quasi-dark dressed state:
\begin{equation}
|\psi_{\mathrm{QD}}\rangle = \frac{1}{\sqrt{N\lambda^2 + g^2}} \left( \sqrt{N}\lambda  |1_{\mathrm{spin}},0_{\mathrm{cat}}\rangle - g  |0_{\mathrm{spin}},1_{\mathrm{cat}}\rangle \right) \otimes |0_{\mathrm{cav}}\rangle,
\label{eq:dark_state}
\end{equation}

\noindent where the stored excitation energy is shared coherently between the battery spin chain and the auxiliary qubit, completely bypassing the lossy cavity sector.

This interference-driven autonomous protection mechanism overcomes several critical bottlenecks inherent to conventional protocols:
\begin{itemize}
    \item Unlike standard dark-state engineering that relies on the stringent tuning of localized individual system-bath couplings~\cite{PhysRevLett.134.180401}, our condition~\eqref{eq:condition} scales with global collective parameters, granting structural robustness against moderate local parameter fluctuations.
    \item Unlike active dynamical decoupling sequences that necessitate fast external time-dependent control fields~~\cite{PhysRevA.107.032615,PhysRevB.111.064503,PhysRevA.111.052622}, our scheme is encoded statically within the system Hamiltonian; the auxiliary qubit operates continuously and autonomously upon initialization.
    \item Unlike traditional models that require globally homogeneous dissipation fields to construct decoherence-free zones~\cite{cyrc-ms34,4hr3-rl2y,msh1-b1cc}, the coherent back-action of the qubit actively reshapes the local relaxation landscape by engineering field nodes.
\end{itemize}

A remarkable numerical confirmation of this mechanism is the constant-energy profile exhibited by the auxiliary qubit (the invariant flat dashed lines in scenario (b) of Figs.~\ref{fig2}--\ref{fig5}). This invariant behavior rigorously demonstrates that the qubit does not act as an active energy source or sink; rather, it behaves as a passive quantum catalyst that alters the dissipation pathways without absorbing or consuming net energy from the battery's work extraction degrees of freedom.

Consequently, the excitation energy injected into the battery becomes dynamically trapped within the quasi-dark subspace governed by Eq.~\eqref{eq:dark_state}. This spectral isolation beautifully accounts for all our numerical findings: the wholesale suppression of initial transient oscillations, the counterintuitive reversal of the adverse $g$-dependent decay trend [Fig.~\ref{fig4}(b)], the formation of long-lived ergotropy plateaus, and the exceptional resilience of the system against both severe cavity decay $\kappa$ and finite thermal noise $T$.

\section{Conclusions}

In this work, we have proposed a catalyst-assisted hybrid open quantum battery model consisting of a spin-chain energy storage unit and an auxiliary coherence-assisted qubit, and established its theoretical framework using the Lindblad master equation. By introducing dual-channel coherent interference, the present scheme achieves autonomous dynamical decoupling, which differs fundamentally from conventional pulsed control strategies.

Numerical results demonstrate that the coherent catalyst effectively suppresses initial transient oscillations of ergotropy and simultaneously enhances the long-time steady-state extractable work. Under variations of internal spin coupling, cavity dissipation, light-matter interaction strength, and environmental temperature, the catalytic structure markedly reduces energy leakage and thermal decoherence. Notably, it maintains favorable performance even under relatively strong dissipation and elevated temperatures, thereby relaxing the operational constraints that limit bare spin-chain quantum batteries.

These findings indicate that interference-assisted auxiliary coherent units offer an effective mechanism for mitigating dissipation-induced degradation in cavity-mediated quantum batteries. Instead of relying on external pulsed driving, the proposed scheme continuously reshapes relaxation dynamics through autonomous coherent interference while the catalyst itself remains energy-conserving. Further investigations incorporating non-Markovian environments and experimental imperfections are still required; nevertheless, the present work provides a promising route toward more robust quantum energy-storage architectures in open environments.

\section*{Author contributions}
S. C. Zhao conceived the idea. N. Y. Zhuang performed the numerical computations and wrote the draft, and S. C. Zhao did the analysis and revised the paper.

\section{Acknowledgment}

This work is supported by the National Natural Science Foundation of China ( Grant Nos. 62065009 and 61565008 ),
Yunnan Fundamental Research Projects, China ( Grant No. 2016FB009 ) and the Foundation for Personnel training projects of Yunnan Province, China ( Grant No. KKSY201207068 ).

\section*{Data Availability Statement}

This manuscript has associated data in a data repository.[Authors' comment: All data included in this manuscript are available upon resonable request by contacting with the corresponding author.]

 \section*{Conflict of Interest}

The authors declare that they have no conflict of interest. This article does not contain any studies with human participants or animals performed by any of the authors. Informed consent was obtained from all individual 
participants included in the study.

\bibliography{references}
\bibliographystyle{apsrev4-2}

\end{document}